\documentclass[amsmath,longbibliography,
reprint,aip,apl]{revtex4-1}

\usepackage{graphicx}
\usepackage{hyperref}

\begin{document}

\title
{Molecular screening effects on exciton-carrier interactions in suspended carbon nanotubes}
\author{T.~Uda}
\affiliation{Nanoscale Quantum Photonics Laboratory, RIKEN Cluster for Pioneering Research, Saitama 351-0198, Japan}
\affiliation{Department of Applied Physics, The University of Tokyo, Tokyo 113-8656, Japan}
\author{S.~Tanaka}
\affiliation{Nanoscale Quantum Photonics Laboratory, RIKEN Cluster for Pioneering Research, Saitama 351-0198, Japan}
\affiliation{Quantum Optoelectronics Research Team, RIKEN Center for Advanced Photonics, Saitama 351-0198, Japan}
\author{Y.~K.~Kato}
\email[Corresponding author. ]{yuichiro.kato@riken.jp}
\affiliation{Nanoscale Quantum Photonics Laboratory, RIKEN Cluster for Pioneering Research, Saitama 351-0198, Japan}
\affiliation{Quantum Optoelectronics Research Team, RIKEN Center for Advanced Photonics, Saitama 351-0198, Japan}

\begin{abstract}
Photoluminescence spectroscopy measurements are performed on suspended carbon nanotubes in a field-effect configuration, and the gate voltage dependence of photoluminescence spectra are compared for the pristine and the molecularly adsorbed states of the nanotubes. We quantify the molecular screening effect on the trion binding energies by determining the energy separation between the bright exciton and the trion emission energies for the two states. The voltage dependence shows narrower voltage regions of constant photoluminescence intensity for the adsorbed states, consistent with a reduction in the electronic bandgap due to screening effects. The charge neutrality points are found to shift after molecular adsorption, which suggests changes in the nanotube chemical potential or the contact metal work function.
\end{abstract}

\maketitle

The Coulomb interactions are enhanced in carbon nanotubes (CNTs) because of the limited screening in one-dimensional systems,\cite{Ogawa:1991, Ando:1997} which result in stable exciton and trion formation even at room temperature.\cite{Wang:2005, Maultzsch:2005, Matsunaga:2011, Yoshida:2016} Their optical properties are sensitive to environmental screening due to their atomically thin nature, and the dielectric environmental effects can explain the variety of emission energies observed from bundles,\cite{Fantini:2004} surfactant wrapped nanotubes,\cite{Weisman:2003} DNA wrapped nanotubes,\cite{Heller:2006} and air-suspended nanotubes.\cite{Lefebvre:2004apa, Ohno:2006prb}

Even for the least screened air-suspended nanotubes, the excitons are known to be affected by adsorbed molecules,\cite{Finnie:2005, Milkie:2005, Homma:2013, Uda:2018, Machiya:2018} and therefore the screening effects can be further reduced when the molecules are desorbed. The spectral shifts in temperature dependence\cite{Finnie:2005} and power dependence measurements\cite{Milkie:2005} have been attributed to heating-induced molecular desorption, while more controlled measurements have shown that water adsorption can explain the spectral changes.\cite{Homma:2013} Taking advantage of the drastic shifts in the emission and absorption spectra, all-optical memory operation\cite{Uda:2018} as well as optical control of coupling between nanobeam cavities and nanotubes\cite{Machiya:2018} have been demonstrated. Excited excitonic states show even more pronounced molecular-screening induced energy shifts because of the small binding energies,\cite{Lefebvre:2008} implying associated changes in the electronic bandgap. Such spectral changes are also expected for trion emission, whereas any modification of the bandgap should influence gate-induced charge accumulation.

Here we investigate the molecular screening effects on trion binding energies and carrier-induced photoluminescence (PL) quenching in suspended CNTs using field-effect transistor (FET) structures. Comparing the gate voltage dependent PL from nanotubes before and after molecular adsorption, we find differences in the emission energies as well as the voltage dependence. The adsorption induced changes in the emission energies indicate that trion binding energies are largely modified by the molecules. Furthermore, the unquenched voltage region for the $E_{11}$ bright exciton PL are narrower for the adsorbed state, which can be explained by a decrease in the electronic bandgap by molecular screening. We also find a shift of the charge neutrality point, which could arise from changes in the chemical potential for the nanotubes or the work function of the contact metals.

\begin{figure}
\includegraphics{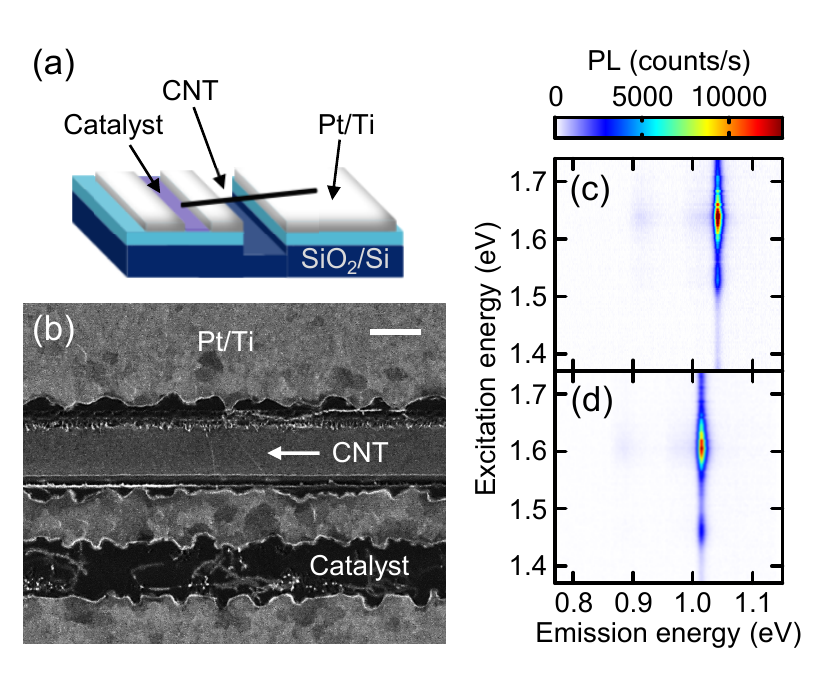}
\caption{
\label{Fig1} (a) A schematic of a CNT FET device. (b) A scanning electron microscope image of a typical device. The scale bar is 1 $\mu$m. (c) and (d) PL excitation maps of a (10,5) nanotube at $V_\text{g}=0$~V in the pristine and the adsorbed states, respectively. The excited $E_{11}$ exciton state shows a more pronounced shift compared to $E_{22}$ resonance.\cite{Lefebvre:2008}
}
\end{figure}

The suspended-nanotube FETs\cite{Yasukochi:2011, Kumamoto:2014, Yoshida:2014, Yoshida:2016, Uda:2016} are fabricated from $p$-doped Si substrates with a resistivity of 0.01-0.02~$\Omega\cdot$cm and a 300-nm-thick oxide layer [Fig.~\ref{Fig1}(a)]. We form trenches using electron beam lithography and dry etching. Another lithography step defines the contact electrodes, and sputtering is used to deposit 1.5~nm Ti and 40~nm Pt. Catalyst areas are patterned on the electrodes with a third lithography step and chemical vapor deposition (CVD) is performed at 800$^\circ$C to grow CNTs over the trenches.\cite{Maruyama:2002} Scanning electron micrograph of a typical device is shown in Fig.~\ref{Fig1}(b). We gate the nanotubes by applying a back gate voltage $V_\text{g}$ to the substrate while grounding the nanotube contacts.

The nanotube devices are characterized with a home-built microspectroscopy system.\cite{Ishii:2015} A wavelength tunable Ti:sapphire laser is used for excitation, and the polarization is rotated using a half-wave plate. An objective lens with a numerical aperture of 0.8 and a working distance of 3.4 mm is used to focus the beam onto a nanotube. We use an excitation power of 30~$\mu$W and the laser polarization is aligned parallel to the tube axis. The same objective lens is used to collect the PL from the nanotube, and the signal is detected by an InGaAs photodiode array attached to a spectrometer. All measurements are carried out at room temperature in dry nitrogen, but we note that the samples have been exposed to air during transfer from the CVD furnace to the measurement system.

Figure~\ref{Fig1}(c) shows a PL excitation map taken within few hours after taking out the sample from the CVD furnace. The emission occurs at the $E_{11}$ energy, while the $E_{22}$ resonance can be observed as a peak in the excitation energy. The $E_{11}$ and $E_{22}$ excitonic resonances appear at higher energies compared to reported values in air ambient,\cite{Lefebvre:2007, Ishii:2015} and match the energies observed for nanotubes without adsorbates.\cite{Chiashi:2008, Homma:2013} We will refer to this state as the “pristine” state, which is metastable and lasts up to 10 hours in air. 

After sufficient time, both $E_{11}$ and $E_{22}$ resonances redshift [Fig.~\ref{Fig1}(d)] to values consistent with tabulated data for air-suspended nanotubes.\cite{Ishii:2015} The shifting of the resonance energies indicate molecular adsorption,\cite{Lefebvre:2008, Chiashi:2008, Homma:2013} and we thus refer to this redshifted state as the “adsorbed” state. Since the molecules can be thermally desorbed\cite{Lefebvre:2008, Xiao:2014} we heat the samples to 400$^\circ$C in Ar gas with 3\% H$_2$ for 20 minutes when we need to convert the nanotubes back to the pristine state.

\begin{figure}
\includegraphics{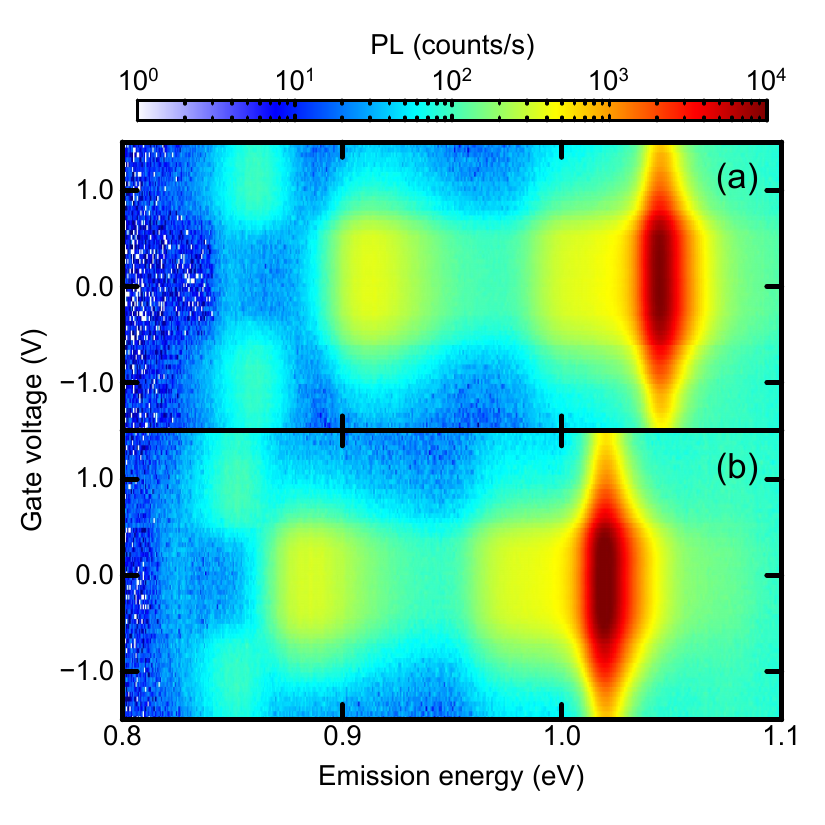}
\caption{
\label{Fig2} (a) and (b) PL spectra as a function of gate voltage for the pristine and the adsorbed states, respectively. The nanotube characterized in Figs.~\ref{Fig1}(c) and \ref{Fig1}(d) is used. The excitation is resonant to the $E_{22}$ excitonic transition energies of 1.64~eV and 1.61~eV for (a) and (b), respectively. 
}
\end{figure}

To investigate the effects of molecular adsorption on the interplay of carriers and excitons, we take PL spectra as a function of gate voltage $V_\text{g}$ for both the pristine and the adsorbed state of this nanotube (Fig.~\ref{Fig2}). Gate-induced quenching of the $E_{11}$  bright exciton emission as well as appearance of the trion emission near 0.85~eV  are observed for both states,\cite{Yoshida:2016} but emission energies are clearly redshifted for the adsorbed state. In addition, there are also subtle differences in the voltage dependence. The pristine state shows a voltage offset of about 150 mV compared to the adsorbed state, while the voltage ranges of $V_\text{g}$ without appreciable PL quenching are slightly different for the two states. 

\begin{figure}
\includegraphics{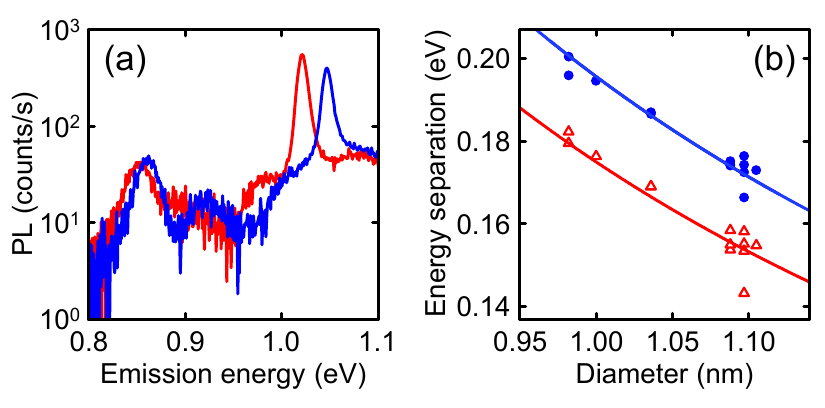}
\caption{
\label{Fig3} (a) PL spectra of the (10,5) tube in the pristine (blue curve) and the adsorbed state (red curve) taken under $V_\text{g}=-1.5$~V. The tube is the same as the one shown in Figs.~\ref{Fig1}(c) and \ref{Fig1}(d). 
(b) Diameter dependence of the energy separation between the exciton and trion states for the pristine states (blue dots) and the adsorbed states (red triangles), respectively. Lines are fits as explained in text. All data are taken with $E_{22}$ excitation.
}
\end{figure}

We first discuss the energy separation between the $E_{11}$ exciton peak and the trion peak. In Fig.~\ref{Fig3}(a), the emission spectra taken at $V_\text{g}=-1.5$~V for the pristine (blue curve) and the adsorbed state (red curve) are plotted. The trion peak shows a shift by 9~meV, which is smaller than the $E_{11}$ exciton peak shift of 24~meV. As the $E_{11}$-trion energy separation $\Delta E$ is the sum of the trion binding energy and the singlet-triplet splitting,\cite{Matsunaga:2011, Santos:2011, Yoshida:2016} it is reasonable that we observe smaller $\Delta E$ for the adsorbed state since screening from the adsorbed molecules should reduce the trion binding energy.

In order to quantify the molecular screening effects on the trion binding energies, we have performed measurements on nanotubes with various chiralities. The obtained values of $\Delta E$ for both pristine and adsorbed states are plotted as a function of the nanotube diameter $d$ in Fig.~\ref{Fig3}(b). It is known that the energy separation between the $E_{11}$ and the trion peak follows the relation $\Delta E=A/d+B/d^2$, where $A$ is the trion binding energy and $B$ is the singlet-triplet splitting.\cite{Matsunaga:2011, Santos:2011, Yoshida:2016} For the adsorbed states, the data points coincide with the previous observation\cite{Yoshida:2016} corresponding to $A=105$~meV$\cdot$nm and $B=70$~meV$\cdot$nm$^2$ (red curve). For the pristine states, energy separations are consistently larger than the adsorbed states by about 20~meV. Assuming that both $A$ and $B$ scale by the same factor $K$ upon molecular desorption, we fit the data by $\Delta E=K(A/d+B/d^2)$. The result is shown as a blue line in Fig.~\ref{Fig3}(b), giving $K=1.12$. This value is comparable to the ratio of 1.10 between the $E_{11}$ exciton binding energies for the pristine state and the adsorbed state,\cite{Lefebvre:2008} and corresponds to a reduction in the trion binding energy caused by the molecular screening of 13~meV for $d=1$~nm tubes. 

\begin{figure}
\includegraphics{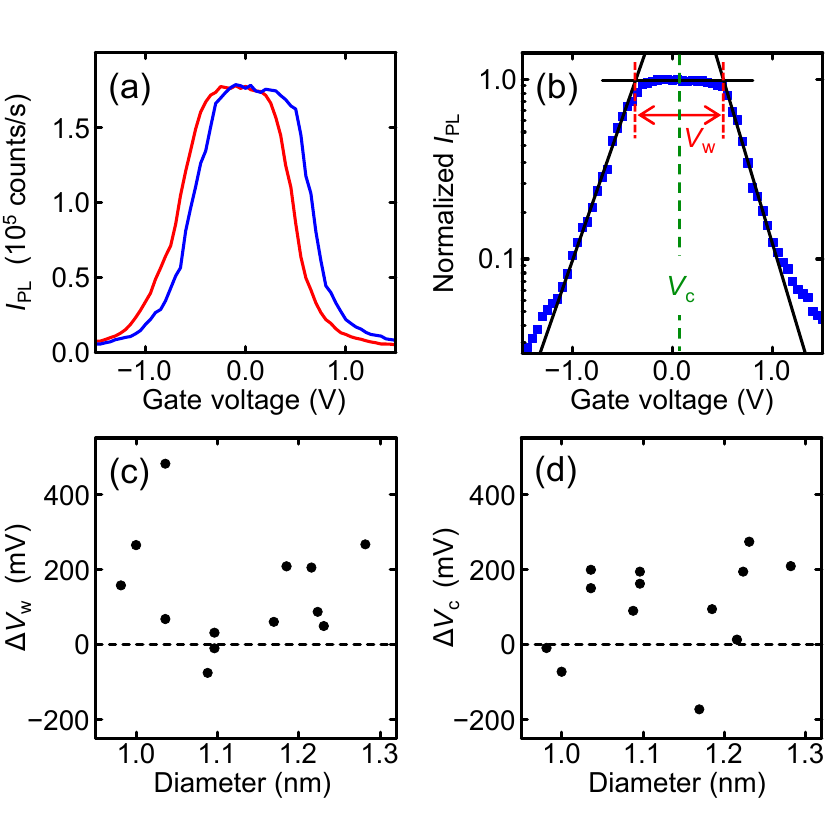}
\caption{
\label{Fig4} (a) Gate voltage dependence of $I_\text{PL}$ for the pristine (blue line) and the adsorbed states (red line). The peak areas are obtained by integrating the PL spectra in Figs.~\ref{Fig2}(a) and ~\ref{Fig2}(b) over 25 pixels in 22-meV-wide spectral windows centered at 1.045~eV and 1.021~eV, respectively. (b) Semi-log plot used to extract $V_\text{w}$ and $V_\text{c}$. The data in (a) for the pristine state are normalized by the maximum value.  (c) and (d) $\Delta V_\text{w}$ and $\Delta V_\text{c}$, respectively, plotted as a function of nanotube diameter. The dashed lines indicate zero. All of the data are taken at $E_{22}$ excitation. 
}
\end{figure}

We now turn our attention to the quenching behavior of the bright exciton emission. For the data sets in Fig.~\ref{Fig2}, we have integrated the PL intensity around the $E_{11}$ emission energies, and the gate voltage dependence of the peak area $I_\text{PL}$  are plotted in Fig.~\ref{Fig4}(a) for the pristine (blue curve) and the adsorbed (red curve) states. Within a voltage range near $V_\text{g}=0.0$~V, the PL intensity is roughly constant, whereas a substantial decrease is observed beyond a threshold for both positive and negative gate voltages. As the quenching occurs due to electrostatically introduced carriers,\cite{Perebeinos:2008, Steiner:2009, Yasukochi:2011} the voltage range with constant PL indicates a region with negligible carrier accumulation, while the gate voltage at the center of the constant PL region corresponds to the charge neutrality point. The voltage offset between the pristine and the adsorbed states is now clearly visible, and the difference in the width of the constant PL region can also be recognized.

To evaluate the differences in a quantitative manner, we extract the voltage range of the constant PL region [Fig.~\ref{Fig4}(b)]. We first draw a line corresponding to the constant PL intensity by averaging all the data points above 95~\% of the maximum intensity. We then take the gate voltage regions where the PL intensity is between 20~\% and 70~\% of the maximum, and fit the data with a line representing exponential quenching with gate voltage. Such fits are performed for both positive and negative gate voltages, and the lines are extrapolated back towards zero gate voltage to obtain the intersections with the constant intensity line. We use the positions of the intersections to determine the width of the voltage range $V_\text{w}$ and the center voltage $V_\text{c}$ for the constant PL intensity region as shown in Fig.~\ref{Fig4}(b).

We have systematically repeated such measurements and analyses on 13 tubes to acquire  $V_\text{w}$ and $V_\text{c}$ for the two states, and we subtract the values for the adsorbed states from those of the pristine states to obtain the differences  $\Delta V_\text{w}$ and $\Delta V_\text{c}$ to identify the molecular screening effects. In Figs.~\ref{Fig4}(c) and \ref{Fig4}(d), we plot $\Delta V_\text{w}$ and $\Delta V_\text{c}$, respectively, as a function of nanotube diameter.  The scatter of the data points are large and there is no apparent dependence on the tube diameter, but we find that there are noticeable differences between the two states. 

The width of the constant PL region is on average larger by about 140~mV for the pristine state compared to the adsorbed state [Fig.~\ref{Fig4}(c)]. This width should be correlated with the electronic bandgap, as it indicates the voltage range for which there is negligible charge accumulation.\cite{Yasukochi:2011, Jiang:2015} It is reasonable that the widths are narrower for the adsorbed state, since the molecular screening of the electron-electron interactions would result in a reduction of the electronic bandgap.\cite{Ando:1997} We note that the observed changes in $\Delta V_\text{w}$ are similar to the values expected from the shifts of excited excitonic states upon molecular desorption.\cite{Lefebvre:2008}

The center voltage, which should indicate the charge neutrality point, shows an average increase of 100~meV for the pristine states. The interpretation for this shift is not straightforward, as our devices have been exposed to air. Although molecular adsorption on the nanotubes can result in slight doping and an associated change in the chemical potential, oxygen is known to cause a change in the work function for contact metals.\cite{Derycke:2002} As we perform the measurements in nitrogen, a decrease in the number of adsorbed oxygen molecules can also cause such a shift. Further study in a more controlled environment is necessary to clarify the origin of the voltage shifts.

In summary, we have investigated molecular screening effects on carrier-exciton interactions in suspended CNTs within field-effect devices by PL spectroscopy. The gate voltage dependence of PL spectra has been measured, and the pristine state is compared to the adsorbed state. Using the energy separation between the bright exciton and the trion emission energies, the effects of molecular screening on trion binding energies have been quantified. For the adsorbed state, we observe narrower voltage regions of constant PL intensity, consistent with a reduction in the electronic bandgap due to screening of the Coulomb interactions by the molecules. We also find that the charge neutrality points shift after molecular adsorption, which may indicate changes in the nanotube chemical potential or the contact metal work function.

\begin{acknowledgments}
Work supported in part by JSPS (KAKENHI JP16H05962) and MEXT (Photon Frontier Network Program, Nanotechnology Platform). T.U. is supported by ALPS and JSPS Research Fellowship. We thank Advanced Manufacturing Support Team at RIKEN for technical assistance.
\end{acknowledgments}

%\bibliography{References}
%merlin.mbs aipnum4-1.bst 2010-07-25 4.21a (PWD, AO, DPC) hacked
%Control: key (0)
%Control: author (8) initials jnrlst
%Control: editor formatted (1) identically to author
%Control: production of article title (0) allowed
%Control: page (1) range
%Control: year (1) truncated
%Control: production of eprint (0) enabled
%

\end{document}